\documentclass[compsoc]{IEEEtran}

\usepackage[utf8]{inputenc}
\usepackage[T1]{fontenc}
\usepackage{mathtools}
\usepackage{times}
\usepackage{a4,latexsym,amsmath,amssymb,graphicx}
\usepackage{hyperref}
\usepackage{caption}
\usepackage[detect-all]{siunitx}
\usepackage{bm}
\DeclarePairedDelimiter\abs{\lvert}{\rvert}%

\providecommand{\keywords}[1]
{
  \small	
  \textbf{\textit{Keywords---}} #1
}

\title{\LARGE \bf
Temporal separation of whale vocalizations from background oceanic noise using a power calculation 
}

\author{\IEEEauthorblockN{Jacques van Wyk\IEEEauthorrefmark{1},
Johan du Preez\IEEEauthorrefmark{2} and Jaco Versfeld\IEEEauthorrefmark{3}}\\
\IEEEauthorblockA{Department of Electrical and Electronic Engineering,
University of Stellenbosch\\ Stellenbosch, South Africa\\
Email: \IEEEauthorrefmark{1}jvanwyk40@gmail.com,
\IEEEauthorrefmark{2}dupreez@sun.ac.za,
\IEEEauthorrefmark{3}djjversfeld@sun.ac.za}}

\begin{document}

\maketitle

\thispagestyle{empty}

\begin{abstract}

The process of analyzing audio signals in search of cetacean vocalizations is in many cases a very arduous task, requiring many complex computations, a plethora of digital processing techniques and the scrutinization of an audio signal with a fine comb to determine where the vocalizations are located. To ease this process, a computationally efficient and noise-resistant method for determining whether an audio segment contains a potential cetacean call is developed here with the help of a robust power calculation for stationary Gaussian noise signals and a recursive method for determining the mean and variance of a given sample frame. The resulting detector is tested on audio recordings containing southern right whale sounds and its performance is compared to a contemporary energy detector and a popular deep learning method. The detector exhibits good performance at moderate-to-high signal-to-noise ratio values. The detector succeeds in being easy to implement, computationally efficient to use and robust enough to accurately detect whale vocalizations in a noisy underwater environment.
\end{abstract}


\section{INTRODUCTION}\label{introduction}

When processing and analyzing digital audio signals, it is often practical to extract only the most useful contents of the complete signal. In speech processing there exist many applications where the automatic differentiation between speech and silence proves useful, such as detecting the silence before and after breaths \cite{FUKUDA201895}. For the purposes of bioacoustic research, such as the study of the acoustic behaviour of cetaceans, researchers may leave a hydrophone array at sea for months before retrieving it. When analyzing these underwater recordings for cetacean calls, where there may be only a few minutes of useful audio data embedded in several hours of underwater noise, the necessity for such a detection method becomes abundantly clear. Numerous methods have been proposed for speech detection using statistical models and maximum likelihood estimation \cite{1510660}\cite{1415230}, as well as detection methods developed for marine use \cite{4449200}\cite{5173498}\cite{Cetacean_Detection}. There are, however, few methods available that are both easy to implement and computationally inexpensive enough to be used in conjunction with more complex detection or classification methods. What is required is a detection method with relatively simple and powerful statistical calculations that provides effective, computationally efficient and robust performance in the presence of background noise without the need for large amounts of labelled training data. Such a detector is developed here with the help of a novel power calculation for Gaussian noise signals \cite{gaussianpower}, to determine the power threshold $P_{\text{thr}}$ and variance threshold $Q_{\text{thr}}$, and a recursive method that is used to calculate the mean and variance \cite{meanvar} of the audio samples. This method should be used as the first step in a detection pipeline, where potential cetacean vocalizations are found within an audio recording containing mostly background noise, after which these potential vocalizations can be verified manually or with a supervised/semi-supervised algorithm, for example a convolutional neural network \cite{CNNBeluga}. It should be emphasized that the proposed technique is not a classification method, simply an energy detector that aims to remove background noise. The goal of this paper is to demonstrate the aforementioned detection method in the context of detecting baleen whale vocalizations and to compare it with contemporary detection methods.

\section{METHOD}\label{sounddet}
The detailed derivations for the power and recursive calculations are shown in appendix A and B respectively. 

\subsection{Average Noise Power Estimation}

The average power of a discrete signal can be expressed as
\begin{equation}
P=\frac{1}{N}\sum_{i=1}^NX_i^2, 
\label{Pavg}
\end{equation}
where $X_i$ is a random variable that represents the instantaneous amplitude of the signal at sample $i$. It is assumed that each $X_i$ is an independent and identically distributed (IID) Gaussian variable. With this a model of the average power of stationary Gaussian noise can be established in the form of a probability density function (pdf) \cite{Joseph1989BayesianSA}. The average power and variance of the estimated power are described by

\begin{equation}
\overline{P} = \sigma_X^2
\label{P_est_avg}
\end{equation} 

\begin{center}
    and
\end{center}

\begin{equation}
\sigma_P^2 = 2\sigma_X^4/N.
\end{equation}

The power of the Gaussian noise signal is assumed to be $\sigma_X^2$ (see Appendix A). This is estimated from the local minima of the signal in Section \ref{stationarity}.

\subsection{Variance in Noise Power Estimation}
The variance of the estimated power can be derived by first noting that the variance of the average power is estimated as

\begin{equation}
Q = \frac{1}{K}\sum_{k=1}^K (P_k - \overline{P})^2
\end{equation}

where $P_k$ is an independent power estimate from Equation \ref{Pavg} and $\overline{P}$ the mean estimated power from Equation \ref{P_est_avg}. The expected value and variance of the variance estimate $Q$ is given by (see Appendix A)

\begin{equation}
\overline{Q}  = \frac{2\sigma_X^4}{N}
\end{equation}

\begin{center}
    and 
\end{center}

\begin{equation}
\sigma_Q^2 =  \frac{8(N+6)\sigma_X^8}{N^3K}, 
\end{equation}

with $K$ the number of independent power estimates and $N$ the number
of samples used in each of those power estimates.

\subsection{Sound detection}
\begin{figure*}
    \centering
    \includegraphics[width=\textwidth]{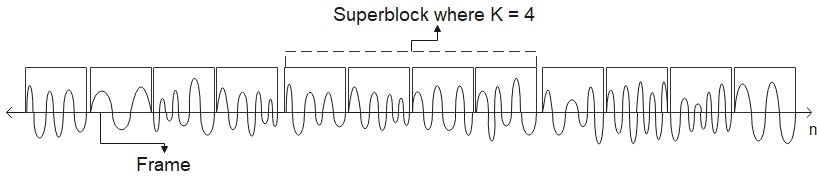}
    \caption{Visual representation of frames and superblocks}
    \label{sample_blocks}
\end{figure*}

The given audio recording can be regarded as a digital signal consisting of $M$ frames that each contain $N$ samples. Each frame corresponds to a power calculation. Adjacent frames can be grouped together into superblocks of $K$ frames. These superblocks correspond to a collection of $K$ power calculations. Figure \ref{sample_blocks} displays an audio signal segmented into frames and superblocks.

A frame is considered silence if the frame $x$ consist only of the Gaussian noise signal $w$: 

\begin{equation*}
    H_0: x = w
\end{equation*}

If a segment contains a whale vocalization $s$ then it is considered a detection:

\begin{equation*}
    H_1: x = w + s,
\end{equation*}

where $x$, $n$ and $s$ are discrete-time vectors \cite{orca}. A suitable threshold model that can discern signals comprised of only noise ($H_0$) from signals that contain possible whale vocalizations ($H_1$) is

\begin{equation}
    H= \begin{cases}
    H_0,& \text{if } P_k < P_{\text{thr}} \text{ and }  Q_k < Q_{\text{thr}}\\
    H_1,              & \text{otherwise}
\end{cases}
\end{equation}
    
where $P_k$ is the mean and $Q_k$ the variance of the power calculations in the superblocks. If a frame or power calculation has a mean or variance below the threshold, then the frame fits within the stationary Gaussian noise model, otherwise it is considered a vocalization. This implies that any signal power or variance above the threshold fits the vocalization model, which is not necessarily true, as the detected signal could be some non-stationary noise signal. To decrease the number of false detections, the value of the thresholds can be increased to allow for more frames to be considered silence. This could in turn cause more of the vocalization frames to be labeled as silence segments. The value of the thresholds are defined as

\begin{equation}
    P_{\text{thr}} = \overline{P}_{\text{est}} + n_{\text{std}}\sigma_P,
\end{equation}

\begin{center}
    and
\end{center}

\begin{equation}
    Q_{\text{thr}} = \overline{Q} + n_{\text{std}}\sigma_Q,
\end{equation}

where $\overline{P}_{\text{est}}$ is the mean of the estimated power, $\sigma_P$ the standard deviation of the estimated power, $\overline{Q}$ is the mean of the estimated superblock variance, $\sigma_Q$ the standard deviation of the estimated superblock variance and $n_{\text{std}}$ is a multiple of the standard deviation. The higher the value of $n_{\text{std}}$, the less likely a signal segment will be labeled as a vocalization. To determine the mean $P_k$ and variance $Q_k$ of each frame and superblock, recursive calculations are used to determine the mean $m_n$

\vspace{0.4cm}

\begin {equation}
m_n =  m_{n-1} + \frac{x_n-m_{n-1}}{n}
\end {equation}

\vspace{0.4cm}

\begin{center}
    and variance $\sigma_n^2$
\end{center}

 \begin{equation}
  \sigma_n^2  = \frac{n-2}{n-1}\sigma_{n-1}^2 + \frac{n}{(n-1)^2}(x_n-m_n)^2,  
\end{equation}

\vspace{0.4cm}

with $n = 2,3 ...$ and $m_1 = x_1$.

\subsection{Stationarity}\label{stationarity}

To determine the value of the thresholds, a rough estimate of the average noise power $\sigma_X^2$ is required. An appropriate estimate could be obtained by taking the minimum amplitude value within a certain timeframe $T$ and then modulating that value to reflect the average noise level within that timeframe. This will result in threshold values that maintain near-constant values over long durations, due to the fact that the background noise is modelled as a stationary random process. Some signals may have noise levels that remain stationary across the whole signal, whilst other signals may have noise levels that change every few seconds or minutes. To find the local minimum $P_{\text{min}}$, an erosion filter \cite{dilEro}\cite{Morph} is applied to the sequence of power calculations $P_k$. The equation

\begin{equation}
    \overline{P}_{\text{est}} = {\alpha}_{f} P_{\text{min}}
\end{equation}

gives us the estimated mean power, where ${\alpha}_{f}$ is the floor modulation coefficient. The floor modulation coefficient should be just large enough that it approximates the noise floor and small enough that crucial low-amplitude signal information isn't lost. A large ${\alpha}_{f}$ value will cause the threshold values to be large as well, which could cause true positives to be disregarded. A good value for ${\alpha}_{f}$ would be $\numrange{1}{10}\;dB$. A dilation filter can also be used to find the local maximum, if the local minimum is too small. The local maximum would then be divided by a coefficient to approximate the noise level. 

It is recommended that some type of impulse filtering algorithm is implemented to remove any unwanted short bursts of non-stationary noise that could be picked up by the detector. A simple way to do this is to remove any detected segment that is shorter than a certain duration.  

Because the detector is based on a stationary Gaussian noise model, it will be referred to as the stationary noise on-line (SNO) detector.

\section{EXPERIMENTS}
\subsection{Energy Detectors}
To test the performance of the SNO detector, audio files containing southern right whale (Eubalaena australis) vocalizations were put through the detector to determine the precision and recall (true positive rate) \cite{prroccurve} of the detector. To gauge the performance of the SNO-detector, RavenPro's band-limited energy detector (BLED)  \cite{2000ASAJ..108.2582M} was used as a comparison tool. The segmentation performance of the detectors were evaluated using Ziółko's \cite{ZIOLKO2015101} fuzzy segmentation scheme to determine how closely a detected segment matches an annotated segment. The precision $P$ and recall $R$ of the detected segments were calculated with

\begin{equation}
    P = \frac{|S \; {\bigcap}_{\mu} \; C|}{S}
\end{equation}

\begin{center}
    and
\end{center}

\begin{equation}
    R = \frac{|S \; {\bigcap}_{\mu} \; C|}{C}
\end{equation}

where $S$ is the segmentation obtained by the detectors, $C$ the correct segmentation (as obtained from an annotation text file or csv file), $|\cdot|$ the duration of the segment and ${\bigcap}_{\mu}$ the intersection between the two sets. \\

The audio files (Microsoft wav format) used to test the detectors are from a collection of underwater recordings obtained by a hydrophone in the coastal waters of False Bay, South Africa. The audio files have a signal-to-noise ratio (SNR) of $20\;dB$.

\begin{center}
    \includegraphics[width=\linewidth, height = 0.21\textheight]{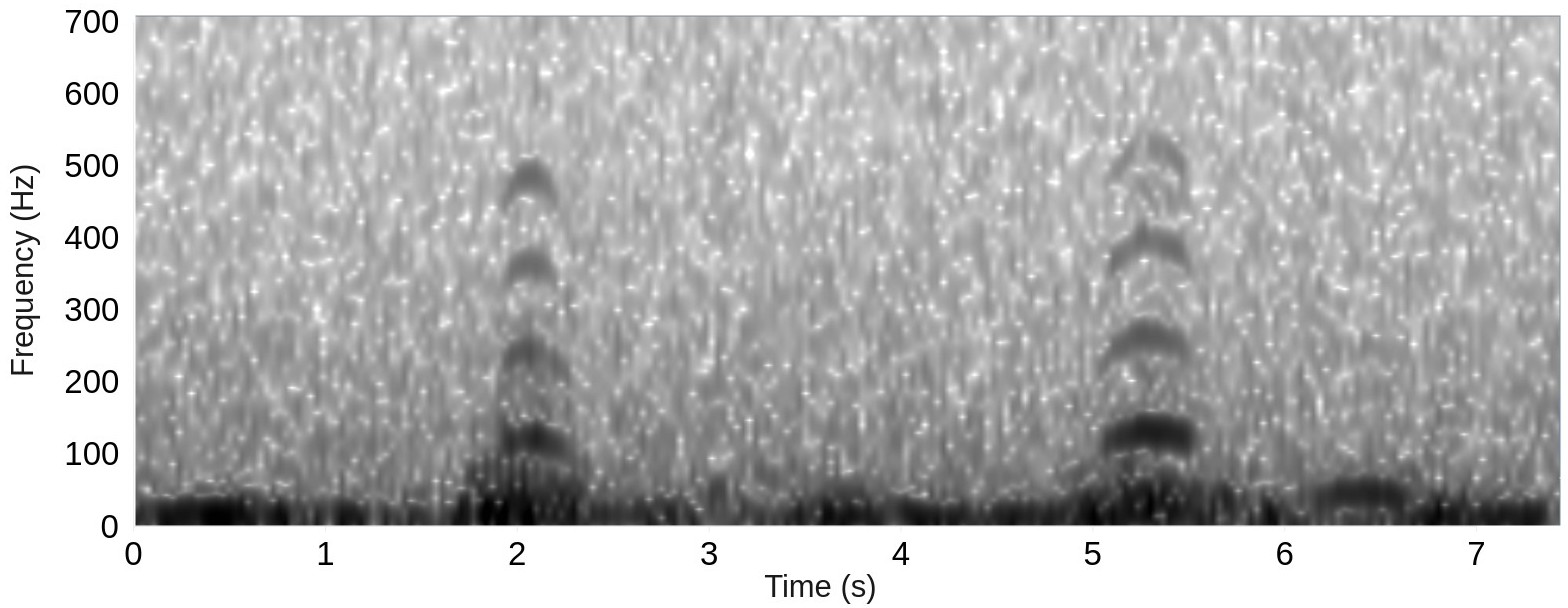}
    \captionof{figure}{Spectrogram of southern right whale sounds}
\end{center}

The audio files were downsampled to $8000\;Hz$ and put through an elliptical filter to yield a passband between $100$ and $700\;Hz$. The code was written in C++ with the help of Stellenbosch's DSP toolkit, which includes the class \textit{onlinesilencedetector.cc} that contains the code to implement the detector as described in Section \ref{sounddet}. 

The parameters of the SNO-detector, such as the frame sizes and target signal lengths, were determined with training sets that were composed of 10\% of the corresponding dataset. The precision and recall curves were obtained by first setting the threshold values to have the highest recall (accept the maximum amount of potential detections), after which the threshold was increased until the maximum precision was reached (the maximum false detections are excluded). For the SNO-detector to function properly it's crucial to select proper values for frame size $N$, superblock size $K$ and stationarity timeframe $T$.  After the parameters were tuned, the detector was run on the test set to produce the precision-recall curves.

\subsection{Deep learning with Resnet-50}
Deep learning has gained popularity over the last few decades as the increase in computational power and available data has made it possible for neural network models to process vast amounts of training data within feasible time spans. A deep learning architecture that has gained considerable interest over the last few years is the deep residual network or RESNET \cite{resnet}. The architecture of this deep learning network uses residual functions and identity shortcut connections that help optimize it's performance without adding any computational complexity, largely solving the degradation problem that very deep models face. Each deep residual network is described by the amount of layers it possesses, i.e. Resnet-34 has 34 layers, Resnet-50 has 50 layers and so on. 

Deep learning models are notorious for requiring large amounts of training data to make accurate predictions, which makes deep learning unappealing for small datasets. Transfer learning assists in the problem of small datasets by using pre-trained weights to aid in the learning process. If the pre-trained weights were used to solve a problem similar to the one the small dataset is trying to solve, then transfer learning can yield favourable results. The feature extractor of the deep learning model can be frozen as not to change these weights or can be made trainable and then the pre-trained weights are used simply for initialization.

To compare the SNO-detector's performance to the performance of a modern convolutional neural network, given meagre training data, a RESNET-50 model was implemented using spectrograms of the audio files as input \cite{CNN_humpback}. The spectrograms were obtained by splitting the audio waveform into 1.2 s non-overlapping windows, then computing the short-time Fourier transform with an overlapping Hann window (50\% overlap) that was 256 samples in length. The resulting spectrogram images had a resolution of $200\times200$ pixels. Because the training dataset is so small and the whale vocalizations so few, we oversampled the training data such that the whale vocalization and noise samples were equal. The oversampling was obtained by re-using whale vocalization samples in the dataset. This prevents the deep learning model from categorizing all the input samples as noise, which is the dominant class in the original dataset. We further augmented the dataset by increasing the overall amount of samples, such that the effective duration of the audio training data was increased to 27 minutes from the original 12 minutes training data. After the model was trained on an equal dataset, we recalibrated the prior odds ratio to compensate for the lower occurance of whale vocalizations (see Appendix C). The actual prior odds ratio is approximately $\text{O}^{'}(W)=\frac{1}{19}$ (where $W$ is the class containing whale vocalizations).
The RESNET-50 was initialized with Imagenet weights and we experimented with frozen and unfrozen feature extractors. The RESNET-50 model was implemented with Tensorflow and the keras API.

\section{RESULTS}\label{results}

The detection performance of the SNO-detector, BLED and RESNET-50 are juxtaposed in Figure \ref{srwcurve}. 
\begin{center}
    \includegraphics[width=\linewidth]{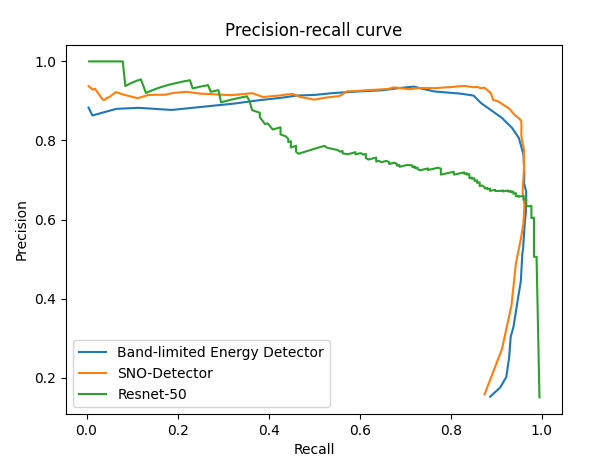}
    \captionof{figure}{Precision-recall curve of southern right whale calls (SNR: 20 dB)}
    \label{srwcurve}
\end{center}

Both energy detectors display satisfactory results when applied to the southern right whale (SRW) audio recordings, which has a high signal-to-noise ratio. From Figure \ref{srwcurve} it is apparent that both detectors perform similarly at low-to-medium threshold values where the recall is high. The initial recall (bottom right of curve) of both energy detectors improve as their thresholds become larger. This happens when the threshold is small enough that the detector groups several whale calls together instead of keeping them separate, which the fuzzy segmentation scheme penalizes. As the threshold becomes larger and the recall decreases, the BLED loses some precision until the threshold becomes high enough that most of the false positives are disregarded. The SNO-detector has precision values that decrease and increase at higher threshold values, which would be caused by the detector obtaining shorter vocalization segments as the quieter parts of the segments fall below the power and variance threshold. The precision would then increase as the threshold becomes larger and disregards more false positives.\\

The best RESNET results were obtained by keeping the layers of the network unfrozen and letting the pre-trained Imagenet weights train on the test data. Initializing the model with these pre-trained weights worked better than if the initial weights were random. The RESNET-model was evaluated by varying the decision threshold of the prediction probabilites.

From Figure \ref{srwcurve} it is apparent that the RESNET-50 performs considerably worse than the SNO-detector given the same amount of training data, except at low recall values (below 30\%), where the RESNET model's precision becomes higher than the SNO-detector's. The RESNET model was given spectrograms of fixed length, which would inevitably lead to multiple shorter whale sounds being grouped together and was accordingly not penalized in it's recall for this behaviour, which is why it's recall does not worsen as the threshold decreases. Neural networks tend to overfit on small datasets, which is why the RESNET-50 struggles to discriminate between whale vocalizations and noise in the test data. Recalibrating the prior odds helped improve the precision of the RESNET predicions by about $7.4\%$, as the system was less eager to award high prediction scores. This also worsened the recall by $1.7\%$, which is an expected and negligible decrease.

Creating spectrograms from audio data and calculating the prediction probability for each spectrogram is computationally expensive. Below are the execution times for converting an audio signal that is 3728 s in duration to 1.2 s spectrograms and having RESNET make predictions for each input spectrogram. 

\begin{center}
\captionof{table}{Execution times for spectrograms and RESNET predictions as applied to an audio signal of 3728 s}
\begin{tabular}{ |p{5cm}|p{3cm}|  }
 
 \hline
 Process & Execution time \\
 \hline \hline
  Convert Audio to Spectrograms & 154.8155 s\\
 \hline
  Make predictions & 209.1615 s\\
 \hline
 \textbf{Total} & \textbf{363.977 s}\\
 \hline

\end{tabular}

\label{executiontimes_spec_resnet}
\end{center}

The deep learning approach takes considerably more time than the SNO-detector, which processes the same audio file within 335 ms. For researchers who could have several weeks of data that need to be processed, a deep learning model would simply take too long.\\

To gauge how fast both energy detectors run on a substantial amount of audio data, a sequence of underwater recordings were put through both detectors and the execution times were measured. The energy detectors were tested on 11 hours of underwater audio and then on 22 hours of underwater audio. The goal of these experiments were simply to measure execution time, not detection performance. The execution times of both detectors are displayed in Table \ref{executiontimes}. The execution times do not take into account the reading of audio files, only the detection times.

\begin{center}
\captionof{table}{Execution times of SNO-detector and BLED}
\begin{tabular}{ |p{2.4cm}||p{2.4cm}|p{2.4cm}|  }
 
 \hline
 Audio Duration &  \multicolumn{2}{c|}{Execution time} \\
 \hline \hline
   & SNO-detector & BLED\\
 \hline
 11 hours & 3.45 s & 24 s\\
 \hline
 22 hours & 6.76 s & 88 s\\
 \hline
\end{tabular}

\label{executiontimes}
\end{center}

From Table \ref{executiontimes} it is apparent that the SNO-detector executes much quicker than the BLED. The difference in execution times could be caused by the difference in computational complexity of the detection algorithms. Another factor is the programming language used to code each method. The SNO-detector was written in C++  while Raven Pro was written in Java. It is expected that any application written in Java, a high-level programming language, would run slower compared to an application written in C++. Since the source code for the BLED is not available, it could not be replicated in C++.\\ 

To gauge the performance of the SNO-detector under low signal-to-noise ratios, the test data was modified with added Gaussian noise that lowered the SNR of the signal to 10 dB. The detection results are displayed in Figure \ref{srwlowsnr}.

\begin{center}
    \includegraphics[width=\linewidth]{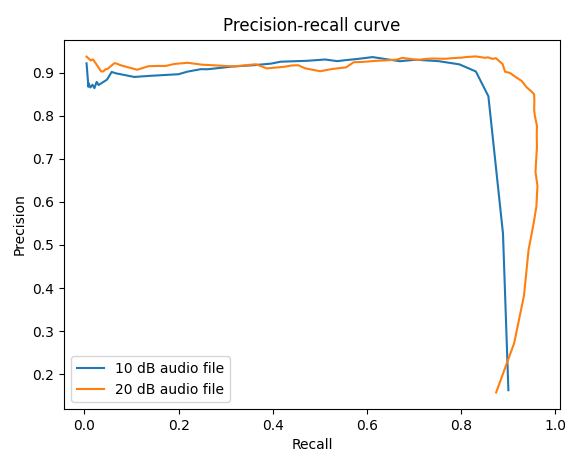}
    \captionof{figure}{Precision and recall curve of SRW calls (SNR: 10 and 20 dB)}
    \label{srwlowsnr}
\end{center}

The SNO-detector performs slightly worse in lower SNR values, as expected, but is robust enough to still deliver good detection results given the increase in Gaussian noise. The largest decrease in performance is at low threshold values, which indicates that the detector either groups several whale calls together or interprets whale calls as noise. At $80\%$ recall the performance of the detector on low SNR values is similar to that at high SNR values.   

\section{DISCUSSION}\label{discussion}
Both the SNO-detector and BLED display good performance given the southern right whale recordings. The SNO-detector maintains a good true positive ratio as the threshold and precision increases, as does the BLED. In the recordings a common source of high frequency noise called "shrimp snapping" was observed. Both energy detectors would filter out the majority of the noise, but there were some loud wideband bursts caused by the shrimp snapping that could be picked up in the whale call bandwidth. These bursts were usually strong enough to be picked up by the detector and not short enough to be disregarded by the impulse filter. This created some false positive results, which is to be expected from an energy detector, as any sound above the power threshold will be regarded as a detection, be it a whale vocalization or noise. The SNO-detector's use of a variance threshold aids in the robustness of the method and helps to ignore a large portion of background noise and recognize most of the whale calls as true positives, although the stronger bursts of non-stationary noise do cause misdetections. From Figure \ref{srwcurve} it is evident that the energy detectors do not reach 100\% precision or recall due to the fuzzy segmentation evaluation, which evaluates how well a detected segment fits its corresponding ground truth segment. The results show that the detected segments don't precisely fit the annotated segments, which is expected. 

It is clear from Figure \ref{srwcurve} that the SNO-detector outperforms the RESNET-50 model, as the deep learning network could not generalize well given the small amount of training data. The SNO-detector also has a much faster execution time in comparison to the RESNET model, processing audio files more than a thousand times faster than it's deep learning counterpart, which makes the SNO-detector much more appealing for researchers who don't have the time and labelled audio data to train a neural network, but quickly want to detect the whale vocalizations that are present in their audio recordings. The SNO-detector could also potentially be used in conjunction with a deep learning model by pre-screening an audio file for whale vocalizations, after which a sufficiently trained neural network can be used to refine the detection results.

The length of $N$ and $K$ must be carefully selected as they aren't initially self-evident and dictate much of the performance of the SNO-detector. A solution to this could be to optimize the sample sizes with regard to some training data before using it for detection. Another consideration is the proper specification of the calls to be detected, as it is impossible to automatically detect every single call in the audio files \cite{listening}. If calls lower than a certain SNR, for example lower than 5 dB, were to be picked up, the precision of the detector would suffer greatly. It is thus important to consider the type of calls to be detected and the overall background noise of the audio recording to avoid negligible results. 
\section{CONCLUSIONS}\label{concl}
When analyzing underwater recordings for cetacean sounds there will almost certainly be noise present, be it from wind, other marine animals, shipping or ice \cite{antarctic}. A capable noise-resistant and effective detection method is required to assist researchers in finding cetacean sounds amidst all this noise, and the stationary noise on-line detector serves this purpose well. The Gaussian power calculations presented here provide a powerful and easily implementable method of modelling the average power and variance of background noise with which an effective and robust detector can be built. In favourable signal-to-noise ratios the SNO-detector can easily differentiate background noise from whale vocalizations, whilst in very noisy environments it occasionally struggles with misdetections and falsely labeling low amplitude signals as silence. 

A potential improvement that can be implemented is in the form of an optimization method to find suitable values for $N$, $K$ and stationarity duration $T$ as well as ${\alpha}_{f}$. If an audio signal has too many bursts of strong noise a correlation-based or deep learning method could be used in conjunction with the noise detector to filter out these strong noise segments after being picked up.

\section*{ACKNOWLEDGMENT}
We would like to thank the National Research Foundation for funding the recording equipment used to capture the audio signals used in our research.

\bibliographystyle{plain}
\bibliography{Bibliography.bib}

\begin{thebibliography}{10}

\bibitem{CNN_humpback}
Ann~N. Allen, Matt Harvey, Lauren Harrell, Aren Jansen, Karlina~P. Merkens,
  Carrie~C. Wall, Julie Cattiau, and Erin~M. Oleson.
\newblock A convolutional neural network for automated detection of humpback
  whale song in a diverse, long-term passive acoustic dataset.
\newblock {\em Frontiers in Marine Science}, 8, 2021.

\bibitem{listening}
Whitlow Au and Marc Lammers.
\newblock {\em Listening in the Ocean}.
\newblock 01 2016.

\bibitem{prroccurve}
Jesse Davis and Mark Goadrich.
\newblock The relationship between precision-recall and roc curves.
\newblock In {\em Proceedings of the 23rd International Conference on Machine
  Learning}, ICML '06, page 233–240, New York, NY, USA, 2006. Association for
  Computing Machinery.

\bibitem{gaussianpower}
Johan du~Preez.
\newblock Power and power variance in a stationary gaussian noise signal.
\newblock personal communication.

\bibitem{meanvar}
Johan du~Preez.
\newblock Recursive estimation of sample mean and variance.
\newblock personal communication, May 2008.

\bibitem{FUKUDA201895}
Takashi Fukuda, Osamu Ichikawa, and Masafumi Nishimura.
\newblock Detecting breathing sounds in realistic japanese telephone
  conversations and its application to automatic speech recognition.
\newblock {\em Speech Communication}, 98:95--103, 2018.

\bibitem{dilEro}
J.Y. Gil and R.~Kimmel.
\newblock Efficient dilation, erosion, opening, and closing algorithms.
\newblock {\em IEEE Transactions on Pattern Analysis and Machine Intelligence},
  24(12):1606--1617, 2002.

\bibitem{resnet}
Kaiming He, Xiangyu Zhang, Shaoqing Ren, and Jian Sun.
\newblock Deep residual learning for image recognition.
\newblock In {\em 2016 IEEE Conference on Computer Vision and Pattern
  Recognition (CVPR)}, pages 770--778, 2016.

\bibitem{Joseph1989BayesianSA}
L.~Joseph and Peter~M. Lee.
\newblock Bayesian statistics: An introduction.
\newblock {\em The American Statistician}, 47:307, 1989.

\bibitem{orca}
Jonas~Philipp Luke, José~G. Marichal-Hernández, Fernando Rosa, and Javier
  Almunia.
\newblock Real time automatic detection of orcinus orca vocalizations in a
  controlled environment.
\newblock {\em Applied Acoustics}, 71(8):771--776, 2010.

\bibitem{Morph}
Petros Maragos.
\newblock Chapter 13 - morphological filtering.
\newblock In Al~Bovik, editor, {\em The Essential Guide to Image Processing},
  pages 293--321. Academic Press, Boston, 2009.

\bibitem{antarctic}
Brian Miller, Naysa Balcazar, Sharon Nieukirk, Emmanuelle Leroy, Meghan Aulich,
  Fannie Shabangu, Bob Dziak, Won~Sang Lee, and Jong Kuk.
\newblock An open access dataset for developing automated detectors of
  antarctic baleen whale sounds and performance evaluation of two commonly used
  detectors.
\newblock {\em Scientific Reports}, 11:806, 01 2021.

\bibitem{2000ASAJ..108.2582M}
Harold {Mills}.
\newblock {Geographically distributed acoustical monitoring of migrating
  birds}.
\newblock {\em Acoustical Society of America Journal}, 108(5):2582, November
  2000.

\bibitem{Peebles}
Peyton~Z. Peebles.
\newblock Chapter 3 - operations on one random variable - expectation.
\newblock In Al~Bovik, editor, {\em Probability, random variables and random
  signal principles}, pages 74--80. McGraw-Hill, Inc, 1980.

\bibitem{1510660}
J.~Ramirez, J.C. Segura, C.~Benitez, L.~Garcia, and A.~Rubio.
\newblock Statistical voice activity detection using a multiple observation
  likelihood ratio test.
\newblock {\em IEEE Signal Processing Letters}, 12(10):689--692, 2005.

\bibitem{1415230}
Jong~Won Shin, Joon-Hyuk Chang, Hwan~Sik Yun, and Nam~Soo Kim.
\newblock Voice activity detection based on generalized gamma distribution.
\newblock In {\em Proceedings. (ICASSP '05). IEEE International Conference on
  Acoustics, Speech, and Signal Processing, 2005.}, volume~1, pages
  I/781--I/784 Vol. 1, 2005.

\bibitem{4449200}
Rustam Stolkin, Sreeram Radhakrishnan, Alexander Sutin, and Rodney Rountree.
\newblock Passive acoustic detection of modulated underwater sounds from
  biological and anthropogenic sources.
\newblock In {\em OCEANS 2007}, pages 1--8, 2007.

\bibitem{5173498}
Ildar~R. Urazghildiiev, Christopher~W. Clark, Timothy~P. Krein, and Susan~E.
  Parks.
\newblock Detection and recognition of north atlantic right whale contact calls
  in the presence of ambient noise.
\newblock {\em IEEE Journal of Oceanic Engineering}, 34(3):358--368, 2009.

\bibitem{Cetacean_Detection}
Ayinde~M. Usman, Olayinka~O. Ogundile, and Daniel J.~J. Versfeld.
\newblock Review of automatic detection and classification techniques for
  cetacean vocalization.
\newblock {\em IEEE Access}, 8:105181--105206, 2020.

\bibitem{CNNBeluga}
Ming Zhong, Manuel Castellote, Rahul Dodhia, Juan Lavista~Ferres, Mandy Keogh,
  and Arial Brewer.
\newblock Beluga whale acoustic signal classification using deep learning
  neural network models.
\newblock {\em The Journal of the Acoustical Society of America},
  147(3):1834--1841, 2020.

\bibitem{ZIOLKO2015101}
Bartosz Ziółko.
\newblock Fuzzy precision and recall measures for audio signals segmentation.
\newblock {\em Fuzzy Sets and Systems}, 279:101--111, 2015.
\newblock Theme: Data, Audio and Image Analysis.

\end{thebibliography}

\pagebreak





\appendices
\section{POWER AND POWER VARIANCE IN A STATIONARY GAUSSIAN NOISE SIGNAL}

\subsection*{A density for the estimated average power of Gaussian noise}

We model a silence signal (with a certain background noise-level) as a
sequence of statistically independent Gaussian random variables $X_i$
with $i=1\ldots N$, each with mean $a_X=0$ and variance $\sigma_x^2$.

Firstly we will derive a density function for the estimated average
power $P$ of $X$.  From this we can determine the mean and variance of
$P$. Let

\begin{equation} 
P=\frac{1}{N}\sum_{i=1}^NX_i^2 \label{powerEstimate}
\end{equation}

be the estimated power. Since the $X_i$'s are random variables, so
will be $P$. We want to find the density $f_P(p)$.

First consider the transformation $Y=X^2$. Then $\frac{dy}{dx}=2x$ and
$x=\pm\sqrt{y}$.

Using basic theory on the transformation of random variables
\cite{Peebles} we find that 

  \begin{align*}
    f_Y(y) & =  \frac{f_X(x)}{\abs{\frac{dy}{dx}}}|_{x=-\sqrt{y}} + \frac{f_X(x)}{\abs{\frac{dy}{dx}}}|_{x=+\sqrt{y}} \nonumber\\
          & =  \frac{f_X(x)}{2\sqrt{y}}|_{x=-\sqrt{y}} + \frac{f_X(x)}{2\sqrt{y}}|_{x=+\sqrt{y}}\\
          & =  \frac{1}{\sqrt{2\pi\sigma_X^2}}y^{\frac{1}{2}-1}e^{\frac{-y}{2\sigma_X^2}}u(y). 
  \end{align*}

Therefore $Y$ is Gamma distributed with parameters $b=\frac{1}{2}$ and
$a=\frac{1}{2\sigma_X^2}$. Its characteristic function is given by:

\begin{equation}
\Phi_Y(\omega) = (\frac{ \frac{1}{2\sigma_X^2} } { \frac{1}{2\sigma_X^2} -j\omega}) ^{\frac{1}{2}}. \nonumber
\end{equation}

Now consider: 
\begin{equation*}
Z=\sum_{i=1}^NY_i \mbox{       with } Y_i \mbox{ IID.} 
\end{equation*}

The characteristic function of $Z$ is now given by:

\begin{align*}
\Phi_Z(\omega) & =  [\Phi_Y(\omega)]^N \\
        & =  (\frac{ \frac{1}{2\sigma_X^2} } { \frac{1}{2\sigma_X^2} -j\omega}) ^{\frac{N}{2}}. 
\end{align*}
  
Therefore $Z$ is also Gamma distributed, but now with parameters $b=\frac{N}{2}$ and
$a=\frac{1}{2\sigma_X^2}$.

The corresponding density function of $Z$ is given by:

\begin{equation}
f_z(z) = \frac{(\frac{1}{2\sigma_X^2})^{N/2}z^{N/2-1}e^{\frac{-z}{2\sigma_X^2}}u(z)}{\Gamma(N/2)}. \nonumber
\end{equation}

The estimated power is calculated as $P=Z/N$. Then $\frac{dp}{dz}=1/N$
and $z=Np$.

\begin{align*}
f_P(p) & =  N\frac{(\frac{1}{2\sigma_X^2})^{N/2}(Np)^{N/2-1}e^{\frac{-Np}{2\sigma_X^2}}u(p)}{\Gamma(N/2)} \\
      & =  \frac{(\frac{N}{2\sigma_X^2})^{N/2}p^{N/2-1}e^{\frac{-Np}{2\sigma_X^2}}u(p)}{\Gamma(N/2)}. 
\end{align*}

Therefore, finally, we can conclude that $P$ is also Gamma
distributed, but now with parameters 

\begin{equation}
b=\frac{N}{2}
\label{gammaB}
\end{equation}

\begin{center}
    and
\end{center}

\begin{equation}
a=\frac{N}{2\sigma_X^2}. 
\label{gammaA}
\end{equation}

From the properties of a Gamma density this also
implies that 

\begin{equation}
\overline{P} = b/a = \sigma_X^2
\label{meanP}
\end{equation} 

\begin{center}
    and
\end{center}

\begin{equation}
\sigma_P^2 = b/a^2 = 2\sigma_X^4/N.
\label{varP}
\end{equation}

\subsection*{The mean and variance of an estimate of the variance of the estimated mean power}

We now estimate the variance $Q$ of the above estimate of the average power $P$. This is given by

\begin{equation}
Q = \frac{1}{K}\sum_{k=1}^K (P_k - \overline{P})^2
\label{estQ}
\end{equation}

with each $P_k$ an independent power estimate such as given by
Eq~\ref{powerEstimate}. From the results of the previous section we know $P_k$ to be Gamma
distributed, and in general we can assume its parameters to be $a$ and
$b$. Its mean is given by $\overline{P}=b/a$ (Eq~\ref{meanP}) and its variance by
$\sigma_P^2 = b/a^2$ (Eq~\ref{varP}). 

The density of $Q$ is somewhat involved, therefore we will rather
focus on directly finding its mean $\overline{Q}$ and variance
$\sigma_Q^2$. First consider:

\begin{equation}
Y = \sum_{k=1}^K (P_k - \overline{P})^2. 
\label{estY}
\end{equation}

Since the $P_k$ terms are statistically independent we can easily find
the mean $\overline{Y}$:

\begin {align}
  E[Y] & =  \sum_{k=1}^KE[(P_k - \overline{P})^2] \nonumber\\
       & =  K\sigma_P^2 \nonumber\\
       & =  Kb/a^2. 
\label{meanY}
\end{align}

The variance of $Y$ can be determined via its moments around the origin:

\begin{align}
  \sigma_Y^2 & =  E[Y^2] -\overline{Y}^2 \nonumber\\
      & =  E[Y^2] - (Kb/a^2)^2.
\label{varY1}
\end{align}

An expression for $E[Y^2]$ is needed:

\begin{align*}
  E[Y^2] & =  E[\sum_{i=1}^K(P_i-\overline{P})^2\sum_{j=1}^K(P_j-\overline{P})^2]  \\
        & =  E[\sum_{i=1}^K(P_i-\overline{P})^4 +\sum_{i=1}^K(P_i-\overline{P})^2\sum_{j=1, j\neq i}^K(P_j-\overline{P})^2]. 
\end{align*}

Taking into account that $P_i$ is independent of $P_j$ with $i\neq j$ it now follows that:
\begin{align*}
  E[Y^2] & = \scriptstyle \sum_{i=1}^KE[(P_i-\overline{P})^4] + \sum_{i=1}^KE[(P_i-\overline{P})^2]\sum_{j=1, j\neq i}^KE[(P_j-\overline{P})^2]  \\
        & =  K\mu_4 + \sum_{i=1}^Kb/a^2\sum_{j=1, j\neq i}^Kb/a^2  \\
        & =  K\mu_4 +(K^2-K)(b/a^2)^2 
\end{align*}

with $\mu_4 = E[(P-\overline{P})^4]$ the 4'th central moment of
$P$. Substitute this back into Eq~\ref{varY1} to find

\begin{align}
  \sigma_Y^2 & =  K\mu_4 + (K^2-K)(b/a^2)^2 - K^2(b/a^2)^2  \nonumber\\
      & =  K[\mu_4-(b/a^2)^2].
\label{varY2}
\end{align}

Using the binomial expression for the power of a sum the required
central moment can be determined as:

\begin{align}
  \mu_4 & =  E[(P-\overline{P})^4] \nonumber\\
     & =  E[\sum_{n=0}^4{4\choose n}P^n(-\overline{P})^{4-n}] \nonumber\\
     & =  E[(-\overline{P})^4 + 4P(-\overline{P})^3 + 6P^2(-\overline{P})^2 + 4P^3(-\overline{P}) + P^4]\nonumber\\
     & =  m_4 - 4m_3\overline{P} + 6m_2(\overline{P})^2-3\overline{P}^4 
\label{mu4}
\end{align}

with $m_n$ the $n$'th moment of $P$ around the origin and
$m_1=\overline{P}$. We therefore still need to find expressions for
the 2nd up to 4th moments of $P$ around the origin. This can be done
via the characteristic function of $P$. In general:

\begin{equation*}
  m_n = E[P^n] = \frac{d^n\Phi_P(\omega)}{j^nd\omega^n}|_{\omega=0}  
\end{equation*}

Therefore:

\begin{align*}
\Phi_P(\omega)                & =  (\frac{a}{a-j\omega})^b = (1-j\omega/a)^{-b}\\ 
\frac{d\Phi_P(\omega)}{d\omega}  & =  \frac{jb}{a}(1-j\omega/a)^{-b-1} \\
\frac{d^2\Phi_P(\omega)}{d\omega^2} & =  \frac{j^2b(b+1)}{a^2}(1-j\omega/a)^{-b-2} \\
\frac{d^3\Phi_P(\omega)}{d\omega^3} & =  \frac{j^3b(b+1)(b+2)}{a^3}(1-j\omega/a)^{-b-3}\\
\frac{d^4\Phi_P(\omega)}{d\omega^4} & =  \frac{j^4b(b+1)(b+2)(b+3)}{a^4}(1-j\omega/a)^{-b-4}. 
\end{align*}

After simplification we therefore find the
required moments around the origin:

\begin{align*}
m_2 & =  \frac{b(b+1)}{a^2} = \frac{b^2+b}{a^2} \\
m_3 & =  \frac{b(b+1)(b+2)}{a^3} = \frac{b^3+3b^2+2b}{a^3} \\
m_4 & =  \frac{b(b+1)(b+2)(b+3)}{a^4} = \frac{b^4+6b^3+11b^2+6b}{a^4}. 
\end{align*}

Substitute this into Eq~\ref{mu4} and simplify to find that  

\begin{equation}
\mu_4 = \frac{3b^2+6b}{a^4}. \nonumber
\end{equation}

Substitute this into Eq~\ref{varY2} and simplify:

\begin{equation}
\sigma_Y^2 = K\frac{2b^2+6b}{a^4}.
\label{varY3}
\end{equation}

We now know $\overline{Y}$ and $\sigma_Y^2$ via Eqs~\ref{meanY} and
\ref{varY3}, and from Eqs~\ref{gammaB} and \ref{gammaA} we know the
appropriate values of $b$ and $a$. From Eqs~\ref{estQ} and \ref{estY}
it follows that $Q=Y/K$. After further simplification (standard
expectation manipulations) we finally get that:

\begin{equation}
\overline{Q} = \frac{\overline{Y}}{K} = b/a^2 = \frac{2\sigma_X^4}{N}
\label{Qmean}
\end{equation}

\begin{center}
    and 
\end{center}

\begin{equation}
\sigma_Q^2 = \frac{\sigma_Y^2}{K^2} = \frac{2b^2+6b}{a^4K} = \frac{8(N+6)\sigma_X^8}{N^3K}, 
\label{Qvar}
\end{equation}

with $K$ the number of independent power estimates and $N$ the number
of samples used in each of those power estimates.

\section{ RECURSIVE ESTIMATION OF SAMPLE MEAN AND VARIANCE}

\subsection*{Standard ML Estimates}

Consider a sequence of samples, namely $ x_1, x_2, ..., x_n,
...$. Note that for the convenience of having the subscript of a
sample also coinciding with the number of samples up to that one, we
chose to start at index $n=1$. The standard unbiased ML estimates for
the mean and variance based on the first $n$ samples are given by:

\begin{equation} \label{mle_mean}
  m_n = \frac{\sum_{i=1}^n x_i}{n}
\end{equation}

\begin{center}
and
\end{center}

\begin{align}
  \sigma_n^2 &=  \frac{\sum_{i=1}^n(x_i-m_n)^2}{n-1} \label{mle_var} \\
	     &=  \frac{\sum_{i=1}^nx_i^2 - \frac{(\sum_{i=1}^n x_i)^2}{n}} {n-1}. \nonumber
\end{align}

\subsection*{Recursive Version of Standard ML Estimates}

We now want to derive expressions to instead do these estimates recursively. We start by rewriting Eq~\ref{mle_mean} as
\begin {align}
m_n &= \frac{\sum_{i=1}^{n-1} x_i}{n} + \frac {x_n}{n} \nonumber \\
    &= \frac{n-1}{n}\frac{\sum_{i=1}^{n-1} x_i}{n-1} + \frac {x_n}{n} \nonumber \\
    &= \frac{n-1}{n} m_{n-1} + \frac {x_n}{n}\nonumber \\
    &= m_{n-1} + \frac{x_n-m_{n-1}}{n}, \label{recur_mean}
\end {align}
with $n = 2,3 ...$ and $m_1 = x_1$.

We can rewrite Eq~\ref{mle_var} as follows:
\begin{align*}
  \sigma_n^2 &= \frac{\sum_{i=1}^{n-1}(x_i-m_n)^2}{n-1} + \frac{(x_n-m_n)^2}{n-1}\nonumber \\
             &= \frac{n-2}{n-1}\frac{\sum_{i=1}^{n-1} [(x_i-m_{n-1}) - (m_n -m_{n-1}) ]^2}{n-2} + \frac{(x_n-m_n)^2}{n-1} \nonumber \\
             &=  \frac{n-2}{n-1}\sigma_{n-1}^2 + \frac{(x_n-m_n)^2}{n-1} + (m_n -m_{n-1})^2.  \nonumber
\end{align*}

For stationary sequences with large $n$ the last term will approach
zero and may (approximately) be omitted. However, this is unnecessary.
We can combine the last two terms by substituting $m_{n-1}$ from
Eq~\ref{recur_mean} and after some simplification get:

\begin{equation}
  \sigma_n^2  = \frac{n-2}{n-1}\sigma_{n-1}^2 + \frac{n}{(n-1)^2}(x_n-m_n)^2.  \label{recur_var} 
\end{equation}

Eqs~\ref{recur_mean} and \ref{recur_var} have an interesting
form. Both express varying coefficient single pole IIR filters.  The
recursion may be initialized by setting $m_1 = x_1$ and $\sigma_1^2 =
0$ and starting computation from $n=2$.

\subsection*{Recursive MAP Estimates}

MAP estimation of Gaussian parameters ultimately reduces to an ML
estimate extended to also include a number $N$ of ``phantom'' samples
that on their own precisely corresponds to the prior mean and variance
values. The above recursive formulations are very easily extended to
include this notion. Instead of initially starting with no prior
knowledge, one now simply starts the recursion at $n=N+1$ with $m_N$
and $\sigma_N^2$ set to the desired prior values.

To understand the role of $N$ it is instructive to remember that it
literally represents a number of imaginary prior observations sampled
from the prior distribution. The higher we select $N$, the more
conservatively our estimate will lean towards the prior
distribution. Or said differently, we need more than $N$ new
observations before their effect in the estimate will start to
override that of the prior model. $N$ is our conservatism parameter
and therefore inversely related to our learning rate.

Also note that nothing prevents us from specifying different values
for $N^{(m)}$ and $N^{(\sigma)}$ for the mean and variance estimates,
thereby effectively giving them different learning rates.

\section{CALIBRATING TWO CLASS CLASSIFIERS}
\subsection{Sigmoidal curves and 2 class LLR scores}

Consider a two class classifier, name the classes $I$ and $T$. The data
to be classified is $\bm{x}$. The posterior probability for class $T$ is
given by Bayes' rule as:

\begin{eqnarray}
  P(T|\bm{x})  &=&  \frac{f(\bm{x}|T)P(T)} {f(\bm{x}|T)P(T)+f(\bm{x}|I)P(I)} \nonumber \\
          &=&  \frac{1} {1+\frac{f(\bm{x}|I)} {f(\bm{x}|T)} \frac{P(I)} {P(T)} }. \label{eq:posterior2class}
\end{eqnarray}

Defining $\text{O}(T) = \frac{P(T)}{P(\overline{T}} = \frac{P(T)}{P(I)}$ as
the ``odds'' of $T$, and $\text{LR}(\bm{x}|T) =
\frac{P(\bm{x}|T)}{P(\bm{x}|\overline{T})}$ as the likelihood ratio of $(\bm{x}|T)$
we can combine the above into:

\begin{eqnarray}
  P(T|\bm{x})  &=&  \frac{1} {1+\text{LR}^{-1}(\bm{x}|T)\,O^{-1}(T) } \nonumber \\
          &=&  \frac{1} {1+e^{-(\text{LLR}(\bm{x}|T)+\text{LO}(T) ) } },  \label{eq:sigmoid}
\end{eqnarray}

with $\text{LLR}(\bm{x}|T) = \log(\text{LR}(\bm{x}|T))$ and $\text{LO}(T) =
\log(\text{O}(T))$.

From this we can see that the sigmoid function gives the relationship
linking the LLR and LO to the the posterior probability. Also note
that the LO is a bias term that simply shifts the sigmoid function to
the left or right.

We can determine LR,O product from the posterior by re-arranging
Eq.~\ref{eq:posterior2class}:

\begin{eqnarray}
  \text{LR}(\bm{x}|T)\,\text{O}(T)  &=&  \frac{P(T|\bm{x})} {1-P(T|\bm{x})}. \label{eq:LROD}
\end{eqnarray}

Since the relationship to the posterior probability is one-to-one,
both contain equivalent information. However, the LR,O product
explicitly states two types of information which are already merged in
the posterior. The LLR measures the evidence we gather from the
observed data ($\bm{x}$ in this case) where-as the LO is concerned with the
prior probability - the information we had before we observed $\bm{x}$.

In the following we will use the LLR,LO combination as \emph{the}
primary summary of the classification information that a system has
available.

\subsection{Recalibrating for a different prior odds ratio}

From the above it is clear that the prior probability (or LO)
influences the posterior probability and therefore also any
calibration resulting in an estimate of the posterior. For which LO
should we therefore train our system? We can, of course, do a separate
calibration for each applicable value of the prior probability, but
this is unnecessarily tedious. Also bear in mind that if either $T$ or
$I$ occur relatively infrequently compared to the other, this data
scarcity will detrimentally affect the training of the
classifier. From this perspective it is useful to train a balanced
classifier with $\text{LO} = 0$ and then later mathematically compensate for
the actual LO value appropriate to a specific application.

The LR,O product (from Eq.~\ref{eq:LROD}) makes it easy to reevaluate
posterior classifier scores using alternative prior probabilities. If
we trained the classsifier with a known prior odds ratio
$\text{O}(T)$, and now want to evaluate it with a different prior odds
ratio $\text{O}^{'}(T)$, we simply multiply the calculated lhs of
Eq.~\ref{eq:LROD} with
$\frac{\text{O}^{'}(T)}{\text{O}(T)}$. Substituting this back into
Eq.~\ref{eq:sigmoid} gives us the re-calibrated posterior as:

\begin{align}
  P(T|\bm{x})  =&  \frac{1} {1+e^{-(\text{LLR}(\bm{x}|T)+\text{LO}(T) + \text{bias})}},  \nonumber \\
    \text{with } \text{bias} =&  \text{LO}^{'}(T)-\text{LO}(T).
    \label{eq:newodds}
\end{align}

Of particular interest is that in both logistic regression, as well as
neural classifiers with sigmoidal activation functions in the final
layer, the expression determining the output is of the form of
Eq.~\ref{eq:sigmoid}. We can therefore easily adapt them for a
different prior log odds by simply modifying their sigmoidal bias
terms.


\end{document}